\documentclass[conference]{IEEEtran}
\IEEEoverridecommandlockouts
\usepackage{cite}
\usepackage{amsmath,amssymb,amsfonts}
\usepackage{graphicx}
\usepackage{textcomp}
\usepackage{xcolor}
\usepackage[pass]{geometry}
\usepackage{algorithm,algpseudocode}
\usepackage{tikz}
\usepackage{tabularx} 
\usepackage{multirow}
\usepackage{array}
\usepackage{microtype}
\usepackage{subcaption}
\usepackage{hyperref}
\hypersetup{draft}  
\usepackage{soul}
\setulcolor{red}
\sethlcolor{yellow}

\def\BibTeX{{\rm B\kern-.05em{\sc i\kern-.025em b}\kern-.08em
    T\kern-.1667em\lower.7ex\hbox{E}\kern-.125emX}}

\makeatletter
\newcommand*\titleheader[1]{\gdef\@titleheader{#1}}
\AtBeginDocument{%
  \let\st@red@title\@title
  \def\@title{%
    \bgroup\normalfont\large\centering\@titleheader\par\egroup
    \vskip1em\st@red@title}
}
\makeatother

\title{Coordinated Management of Processor Configuration and Cache Partitioning to Optimize Energy under QoS Constraints}
\titleheader{\textit{Submitted to: 34th IEEE International Parallel \& Distributed Processing Symposium (IPDPS2020)}}

\author{
\IEEEauthorblockN{Mehrzad Nejat, Madhavan Manivannan, Miquel Peric{\`a}s, Per Stenstr{\"o}m}
\IEEEauthorblockA{
    \textit{Department of Computer Science and Engineering, }
    \textit{Chalmers University of Technology}\\
    \textit{Email: \{nejatm, madhavan, miquelp, per.stenstrom\}@chalmers.se}\\
    }
}

\begin{document}

\maketitle
\thispagestyle{plain}
\pagestyle{plain}
\vspace{-20pt}
\begin{abstract}
An effective way to improve energy efficiency is to throttle hardware resources to meet a certain performance target, specified as a QoS constraint, associated with all applications running on a multicore system.

    Prior art has proposed resource management (RM) frameworks in which the share of the last-level cache (LLC) assigned to each processor and the voltage-frequency (VF) setting for each processor is managed in a coordinated fashion to reduce energy. A drawback of such a scheme is that, while one core gives up LLC resources for another core, the performance drop must be compensated by a higher VF setting which leads to a quadratic increase in energy consumption. By allowing each core to be adapted to exploit instruction and memory-level parallelism (ILP/MLP), substantially higher energy savings are enabled.

This paper proposes a coordinated RM for LLC partitioning, processor adaptation, and per-core VF scaling. A first contribution is a systematic study of the resource trade-offs enabled when trading between the three classes of resources in a coordinated fashion. A second contribution is a new RM framework that utilizes these trade-offs to save more energy. Finally, a challenge  to accurately model the impact of resource throttling on performance is to predict the amount of MLP with high accuracy. To this end, the paper contributes with a mechanism that estimates the effect of MLP over different processor configurations and LLC allocations. Overall, we show that  up to 18\% of energy, and on average 10\%, can be saved using the proposed scheme.

\end{abstract}


\section{Introduction} \label{sec_intro}

Improving energy efficiency of processors is an important goal. An effective way to achieve this goal is to dynamically throttle hardware resources to meet a certain performance target, thereby reducing energy consumption. To realize this, applications must be associated with Quality of Service (QoS) targets that specify performance constraints. 


A Resource Manager (RM) implemented as part of a runtime system is typically used for this purpose. In prior art, RMs have been proposed to control Dynamic Voltage Frequency Scaling (DVFS) \cite{MOGHADDAM2017,Suh2015DynamicProcessors,Choi2002Frame-basedDecoder} and combinations of DVFS and dynamic adaptation of  processor resources (henceforth also referred to as core configuration) \cite{Pothukuchi2016UsingArchitectures,Hughes2001SavingApplications} to meet an application's QoS target while reducing energy consumption. These resources are private to a single application. But, other resources, e.g. the Last-Level Cache (LLC), are shared among different applications in a multi-core system. These schemes do not exploit the resource trade-offs between applications that share a LLC. 

Prior work, such as \cite{Lo2015Heracles:Scale}, considers the partitioning of LLC to respect QoS for a single application, while utilizing the remaining space for best effort jobs. But, the workloads are shifting toward supporting multiple applications with QoS constraints \cite{PARTIES2019}. Hence it is important for RMs to exploit resource tradeoffs across such applications. Consider a workload that is a mix of memory intensive applications. It is possible to reduce system energy substantially by tuning the cache allocations to minimize the number of memory accesses. However, the scope for optimizing cache resources is limited when respecting the performance constraints of every application. To alleviate this limitation, Nejat \textit{et al.} \cite{Nejat2019} proposed an RM that throttles both LLC resources and per-core DVFS in a coordinated fashion. With that scheme, a first application can give up portions of its LLC share to a second one that benefits more from it by having the first application compensate for its performance loss at a slightly higher voltage-frequency (VF) level. This allows the second application to meet its QoS target at a reduced VF. This RM is designed to assess a quite large configuration space at runtime with negligible overhead to find the most energy-efficient resource allocations.


The scope of energy reduction using DVFS and cache partitioning alone is however limited. When an application with a reduced LLC share attempts to compensate the performance impact of additional cache misses by increasing core VF, it imposes a quadratic energy cost.  Alternatively, if the RM can also throttle resources in the core micro-architecture, it can achieve the required performance improvement by exploiting more instruction and memory-level parallelism (ILP and MLP, respectively). The energy cost of this action is substantially lower because of the often linear relation between core size and energy. Furthermore, with the improvement in MLP, application performance becomes less sensitive to the number of cache misses. This will relax the constraints on LLC partitioning and allow for more efficient distributions of LLC shares. 

In this paper, we propose an online RM that orchestrates the tuning of micro-architecture and VF settings of each core with partitioning of the shared LLC. It dynamically evaluates different resource settings to minimize system energy under QoS constraints. Prior to implementation of this RM, an understanding of the complex energy and performance trade-offs between the aforementioned resources is required. We present a systematic analysis of these trade-offs in a diverse range of workload scenarios. This analysis identifies the scenarios where the proposed scheme achieves significant improvements and where it has limited effect.



An online RM requires a means to estimate the effect of resource allocation decisions on performance and energy for each application. Modeling the complex interactions between these resources and their impact on performance and energy is a major challenge. To simplify this problem, we use a core architecture that can be configured to a limited number of sizes  with a balanced pipeline. This is done by deactivating sections of core components:  reducing the issue width,  the load/store queue size, the reorder buffer size, the number of reservation stations, and the number of active functional units. Prior work has shown that such reconfiguration capabilities can be implemented with relatively low overhead \cite{ElasticCore2018, DynTunProc}.

We utilize analytical models to estimate the effect of changing the core size along with DVFS and LLC allocation, on both performance and energy. However, the accuracy of the online performance model strongly depends on an estimation of memory access time based on MLP. Therefore, we propose a low cost scheme that builds on the Auxiliary Tag Directory (ATD) \cite{Qureshi2006Utility-BasedCaches} mechanism. ATD is an on-line technique to estimate the number of cache misses for different cache sizes. The proposed scheme uses a heuristic method to count only the leading misses in a group of overlapping memory accesses, for each possible core size.


We make the following contributions: 
\begin{enumerate}
    \item A systematic analysis to understand the core and cache resource trade-offs for different workload scenarios. The analysis is based on categorization of applications according to level of ILP/MLP and cache sensitivity. 
    \item An online RM that exploits the trade-offs introduced by leveraging ILP/MLP when orchestrating the control of core configurations, VF settings and  LLC partitioning under QoS constraints. 
    \item A hardware design for online estimation of MLP across a range of core configurations and LLC allocations. This enables fast and accurate prediction of performance and energy as a function of resource settings at run time. 
    \item An evaluation of the energy savings by the proposed RM. We show that our proposed scheme can save up to 18\%, 10\% on average, with a low likelihood of violating QoS, compared to a baseline system setting with evenly distributed shared resources. 
\end{enumerate}

The rest of the paper is organized as follows. Section \ref{sec_motivation} provides motivation for this work. The proposed scheme is described in Section \ref{sec_proposed}. Section \ref{sec_exper_method} and \ref{sec_exper_res} present the experimental methodology and evaluation results, respectively. The related work is discussed in Section \ref{sec_related} and, finally, Section \ref{sec_conclude} concludes the paper. 

\section{Background and Motivation} \label{sec_motivation}

Our baseline system model is a conventional multicore system running a multiprogrammed workload where each application is associated with a QoS target expressed as a performance constraint. The baseline has a fixed setting for the resources considered in this paper: even distribution of LLC shares across cores, and a default, mid-range setting of VF and core size. 
This setting meets the QoS targets of all applications in the multiprogrammed workload.


\begin{figure}[t]
    \centering
    \includegraphics[width=0.95\linewidth]{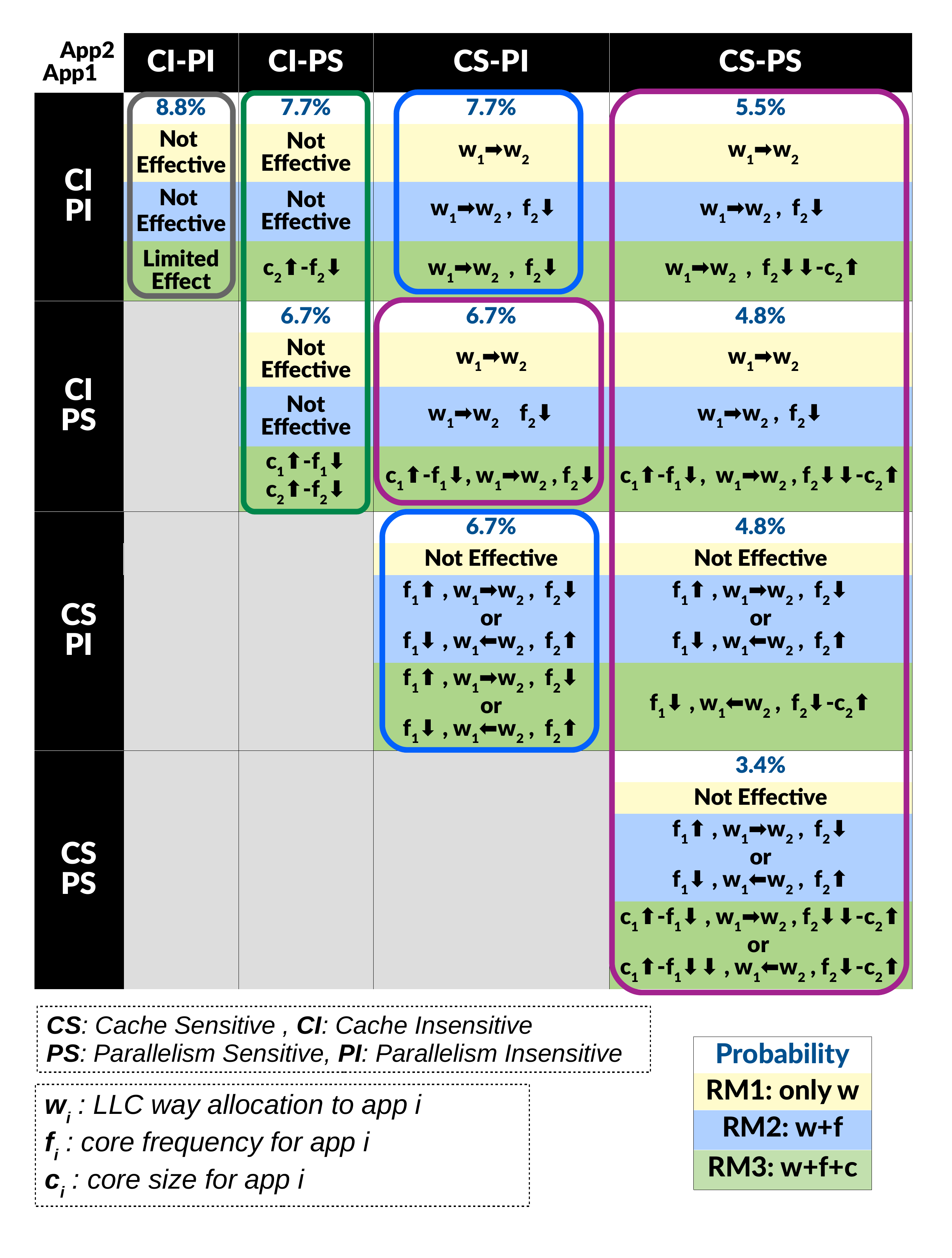}
    \vspace{-10pt}
    \caption{Potential resource trade-offs in different workload mixes.}
    \label{fig_scenarios_table}
    \vspace{-10pt}
\end{figure}

To reason about the configuration space for resource settings, we characterize applications concerning their need for more cache space -- \emph{cache sensitivity} -- and more micro-architectural core resources \emph{parallelism sensitivity}:

\begin{enumerate}
    \item \textit{Cache Sensitivity/Insensitivity -- CS/CI}: An application is considered to be Cache Sensitive (CS) if a change to the baseline LLC allocation leads to a certain variation in Misses Per Kilo Instructions (MPKI). Otherwise, it is Cache Insensitive (CI). 
    \item \textit{Parallelism Sensitivity/Insensitivity -- PS/PI}: An application is considered Parallelism Sensitive (PS) if a change in core size leads to a certain variation in instruction-level or memory-level parallelism (ILP or MLP, respectively). Otherwise it is Parallelism Insensitive (PI). 
\end{enumerate}


To reason about the tradeoffs between resources, we consider workload mixes comprising two applications, where each application belongs to any of the four categories defined above. All possible mixes are depicted in Figure~\ref{fig_scenarios_table}. We have omitted the lower triangular part of the table as it is symmetric with respect to the upper triangular part. For each mix in the table, a probability is calculated based on the number of SPEC 2006 benchmark applications in each category (for details, refer to Section~\ref{sec_exper_method}). For example, 5 out of 27 benchmarks belong to CS-PS. Thus, a mix of two such applications has $3.4\%$ probability. We consider three distinct RMs. RM1 performs only LLC partitioning. RM2 partitions LLC in coordination with per-core VF scaling according to \cite{Nejat2019}. In addition to this, RM3 can also re-configure the size of the cores.


\begin{figure}[t]
    \centering
    \includegraphics[width=0.85\linewidth]{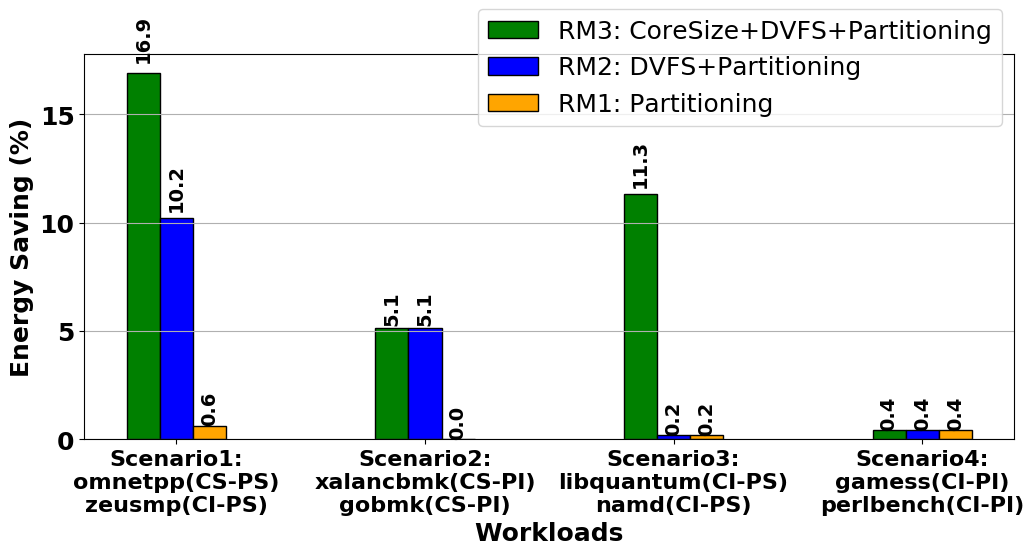}
    \vspace{-7pt}
    \caption{Simulation results for two core workload scenarios.}
    \label{fig_motiv_results}
    \vspace{-10pt}
\end{figure}

Each cell depicts the potential resource trade-offs to save energy without any performance degradation in relation to the assumed baseline system setting. $\text{f}_i\uparrow$ and $\text{f}_i\downarrow$ is used to denote an increase and decrease in the VF of core i. Similarly, the increment and decrement in core size is expressed by $\text{c}_i\uparrow$ and $\text{c}_i\downarrow$, respectively. $\text{w}_i\rightarrow \text{w}_j$ denotes a redistribution of LLC shares from application i to application j. For example, in the mix comprising two CS-PS applications, RM1 cannot change the baseline partitioning without violating QoS of one application. RM2, on the other hand, can increase the LLC share of the application that experiences more LLC misses and can reduce its VF to save core energy. However, VF of the other core must be increased to satisfy QoS. In contrast, RM3 can provide the most savings by increasing the core sizes, thereby exploiting more ILP/MLP. Consequently, the core with an increased LLC share can further reduce its VF, denoted $\downarrow\downarrow$, while the other core can also operate at a lower VF. In essence, this can potentially make the performance less sensitive to the number of cache misses and allows a more efficient distribution of LLC shares to reduce the number of  memory accesses.

Figure~\ref{fig_scenarios_table} shows four distinct scenarios regarding potential energy savings which are highlighted using bounded rectangles. Scenario~1, in purple, indicates a case where the proposed resource manager, i.e. RM3, is expected to provide additional energy improvement compared to prior art (RM2).  Scenario~2, in blue, indicates a case where both RMs are comparable. Scenario~3, in green, indicates where only RM3 is effective. Finally, Scenario~4, in grey, is used for the case where limited/no energy savings are expected by either RM. We observe that RM3 is more effective in 12 out of 16 mixes with a collective probability of 70\%. It should be noted that a reduction of core size is not presented in this analysis because of the following reason. According to our evaluations, there are only few cases where selecting the smallest core size leads to considerable energy saving without violating QoS.

To evaluate the energy savings, Figure~\ref{fig_motiv_results} shows the simulation results for specific two-core workloads corresponding to each scenario. The simulations are performed using the methodology explained in Section~\ref{sec_exper_method} with perfect assumptions regarding modeling accuracy and overheads. The figure shows 70\% higher energy saving with RM3 compared to RM2 in Scenario~1. Both RMs save similar amount (5\%) of energy in Scenario~2. Only RM3 is effective in Scenario~3 and saves 11\% of system energy. In Scenario~4, all the three RMs are ineffective.

\section{Proposed Scheme} \label{sec_proposed}
This section presents the proposed RM scheme. It builds on an RM framework proposed by Nejat et al. \cite{Nejat2019} that meets QoS targets of each application in a multiprogrammed workload and is used as a baseline in this study. It does so by controlling LLC resources and VF scaling of individual cores in a coordinated fashion with the goal of reducing energy consumption. This framework is introduced in Section \ref{framework}.

This paper introduces the notion of core adaptation as an important concept added to the baseline framework. To incorporate it calls for new concepts to accurately model the impact of memory-level parallelism on performance and energy. We present our proposed solutions in Sections \ref{subsec_perf_model} and \ref{subsec_energy_model}, respectively. 


\begin{figure}[t]
    \centering
    \includegraphics[width=\linewidth]{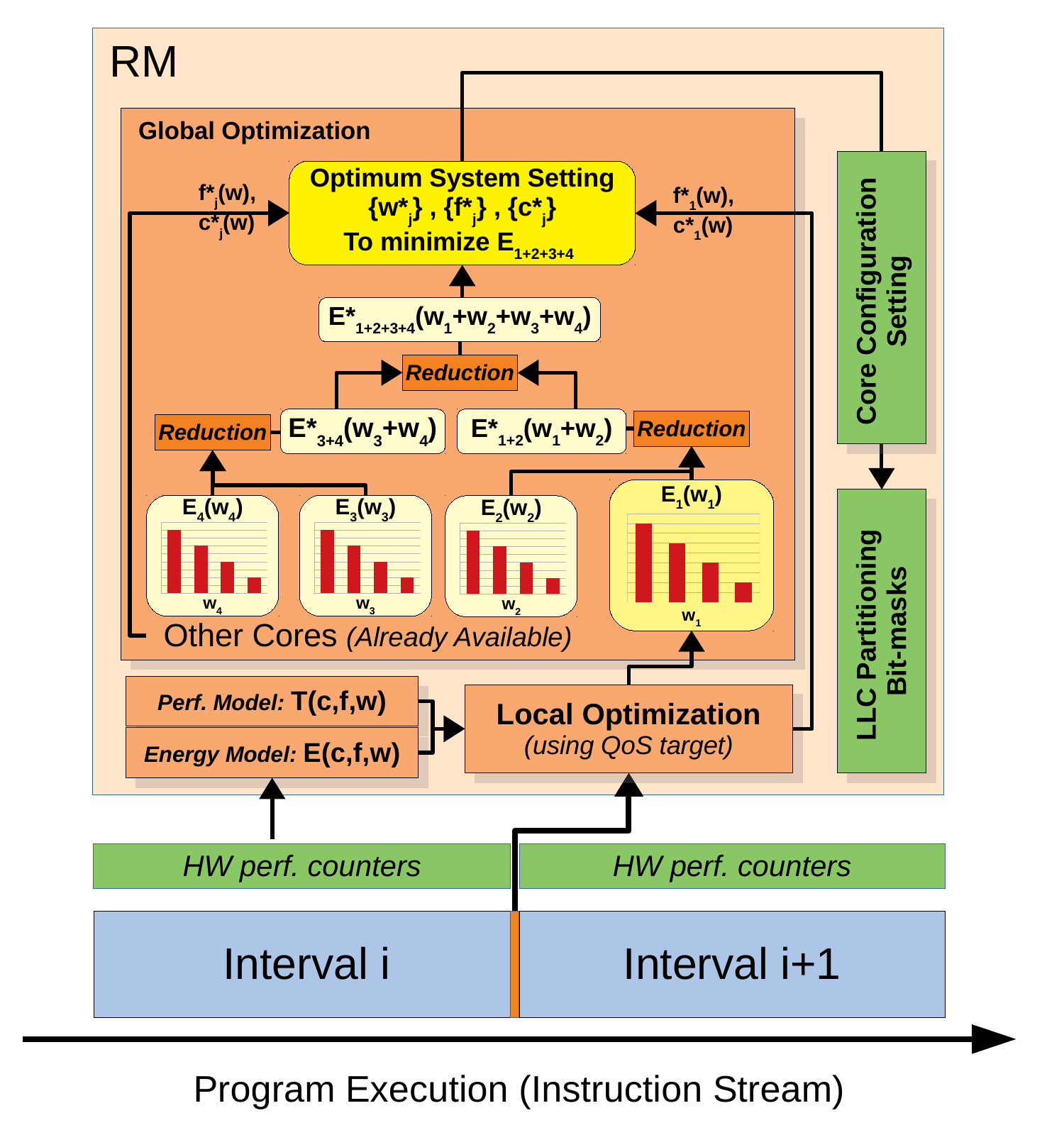}
    \vspace{-20pt}
    \caption{An overview of the proposed RM.}
    \label{fig_rm_scheme}
    \vspace{-10pt}
\end{figure}


\subsection{Baseline RM Framework}
\label{framework}
 An overview of the RM framework is illustrated in Figure~\ref{fig_rm_scheme}. In the baseline framework, the RM is invoked in each core every time a fixed number of instructions has been executed, called interval. It starts with finding a minimum frequency (f$^*$) for each possible allocation of LLC ways (w) that satisfies QoS i.e. leads to an execution time less or equal to that of the baseline setting. It uses simple analytical models to predict the execution time of the upcoming interval (i+1) for different cache allocations and VF settings including the baseline. The models use the statistics collected from hardware performance counters as well as from the Auxiliary Tag Directory (ATD) \cite{Qureshi2006Utility-BasedCaches} over the past interval (i). ATD generates the number of cache misses as a function of w during runtime (more details in Section~\ref{subsec_perf_model}). However, it does not provide any estimation of MLP. Therefore, the performance model in the baseline framework divides the number of misses by the average MLP measured over the past interval, for every w. This assumption of constant MLP leads to modeling error which is evaluated in Section~\ref{sec_exper_res}.  
 
 In the next step of the local optimization, the energy model is used for each pair of w and $\text{f}^*$(w) to generate an energy curve ($\text{E}_1(\text{w}_1)$ in Figure~\ref{fig_rm_scheme}). This curve is passed to the global optimization algorithm that already contains the previously generated curves for other cores. The algorithm recursively reduces pairs of energy curves until it finds an optimum LLC distribution $\{\text{w}^*_j\}$ for each core $j$ that minimizes $\sum_{j}\text{E}_j(\text{w}_j)$ while respecting the resource constraint $\sum_{j}\text{w}_j=\text{A}$; where A is the associativity of LLC. Finally, $\text{w}^*_j$ and the corresponding f$^*_j$(w$^*_j$) setting is applied to each core $j$.
 
 This scheme has two important advantages. First, it has polynomial time complexity with respect to the number of cores. Second, the interface between local and global optimization is an energy curve that can be generated with different combinations of local resources without affecting the global optimization.

\subsection{Overview of the Proposed Core Adaptation RM}
The simple local optimization used in the previous scheme cannot be easily extended to support core adaptation. The assumption of constant MLP leads to a significant modeling error when changing the core size. This results in large QoS violations as well as reduced energy savings (see Section~\ref{sec_exper_res}).  Therefore, in this work, we propose a new modeling framework to address this problem as explained in the rest of this section. The proposed local optimization algorithm utilizes this framework to generate a c*(w) and f*(w) along with an energy curve. These functions determine the core size and frequency that satisfy QoS with minimum energy for each cache allocation w. The global optimization performs the same recursive procedure explained earlier to find the optimum LLC distribution $\{\text{w}^*_j\}$. Finally, the new system setting is applied according to this distribution and the corresponding $\text{c}^*_j$(w$^*_j$) and $\text{f}^*_j$(w$^*_j$) for each core $j$. 

\subsection{Performance Modeling} \label{subsec_perf_model}
The proposed RM requires a performance model to predict the execution time for a range of cache allocations and core configurations. The model should be fast enough to impose negligible runtime overhead and provide enough accuracy to minimize the number of QoS violations. Furthermore, it should not depend on prior knowledge about the applications to make this a general solution. 

ATD \cite{Qureshi2006Utility-BasedCaches} is a commonly used technique for online generation of cache-miss profiles. The address of every memory request that access the main cache is captured by the ATD. It emulates the operation of the main tag directory to detect if the access hits in any recency position or misses. The number of misses is predicted for any allocation of w cache ways by summing up the number of hits in recency positions greater than w with the number of ATD misses. 


The total number of cache misses cannot be used to achieve an accurate estimation of memory stall time due to MLP. For every group of overlapping memory access, only the stall time of the \textit{Leading Miss} (LM) must be counted \cite{LeadingLoads2014, Miftakhutdinov2012, spiliopoulos2016unified}. However, to the best of our knowledge, no online solution has been proposed to provide such statistic over different core configurations and LLC allocations. Hence, we propose a low cost extension to the ATD to address this problem.

\begin{figure}[t]
    \centering
    \includegraphics[width=\linewidth]{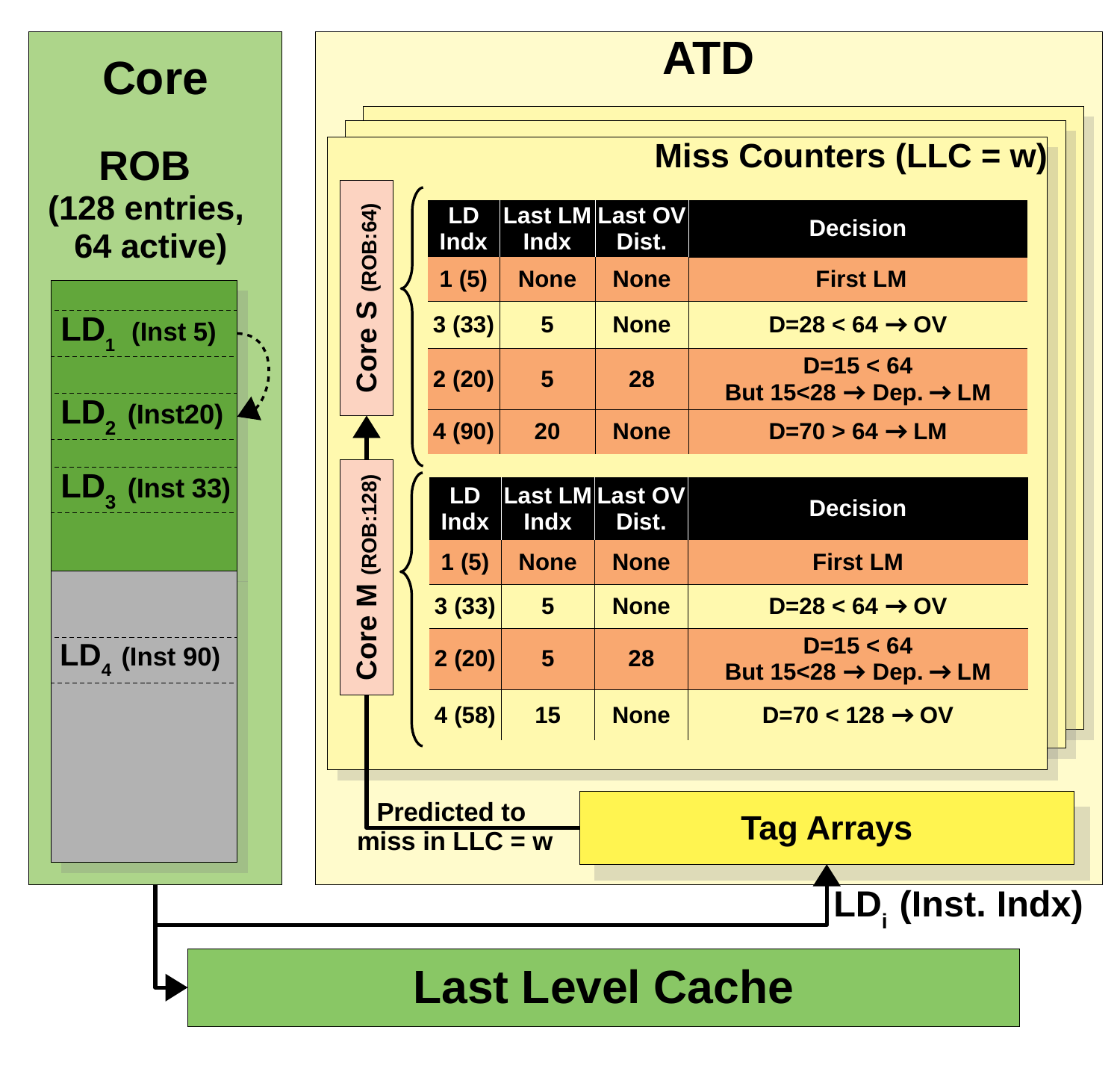}
    \vspace{-20pt}
    \caption{The proposed extension to ATD for MLP estimation.}
    \label{fig_atd_extn}
    \vspace{-10pt}
\end{figure}

In our solution, a new miss counter is added for every core size and every cache allocation. We use the simple example depicted in Figure~\ref{fig_atd_extn} to explain the behaviour of these counters. In the example, the core can be configured to S(mall) or M(edium) with a reorder buffer (ROB) size of 64 and 128 entries, respectively. There are four loads in the instruction stream that will access the LLC and ATD. All these accesses are predicted to miss in LLC allocation w. Each LLC access that goes to the ATD carries a few more bits that indicate the instruction index. This index represents the location of each instruction over a fixed instruction window. In this case, we have pessimistically used a window size equal to four times the maximum size of the ROB which requires 10 bits. Our heuristic to detect overlapping (OV) misses for each core size is based on the following criteria. A particular memory access is counted as OV if:  
\begin{enumerate}
    \item Its distance to the last LM  is less than the ROB size. 
    \item It does not have a data dependency with the last LM. 
\end{enumerate}
The first criterion can be simply evaluated by storing the instruction index of each LM and comparing it to that of the following loads. But, evaluating the second criterion is more complicated. To address this issue, we make a simplifying assumption: If load instructions arrive out of order at the ATD, it is likely due to a data dependency to an earlier load. In this example, $\text{LD}_2$ has to wait for the data collected by $\text{LD}_1$. Thus, the independent $\text{LD}_3$ can bypass $\text{LD}_2$ and arrive earlier at the ATD. To capture this behaviour, another register is needed to store the distance of last OV to the last LM. Therefore, in the miss counter for the S core in Figure~\ref{fig_atd_extn}, $\text{LD}_1$ is counted as LM since it is the first load. Thus, its instruction index is stored in \textit{``Last LM Indx''} register. $\text{LD}_3$ is ignored as its distance to last LM is smaller than ROB size and there is no prior OV distance. The distance of $\text{LD}_3$ to the last LM is stored in \textit{``Last OV Dist.''} register. Next, $\text{LD}_2$ arrives with a distance to last LM less than the ROB size. But, it is less than the previous OV distance indicating an out-of-order arrival. Thus, it is counted as a new LM due to data dependency with previous LM. The two registers are updated accordingly. Finally, $\text{LD}_4$ is also counted as a new LM since its distance to the last LM is greater than the ROB size. In contrast, the miss counter for the M core counts $\text{LD}_4$ as overlapping. Because its distance to the last LL is smaller than the ROB size and no dependency is detected for this load. 

Using the proposed hardware technique, we develop the performance model to estimate the execution time of the upcoming interval $i+1$ for a target setting (c,f,w) as: 

\begin{align}  \label{eq_t_model}
    \text{T}_{i+1}(\text{c},\text{f},\text{w}) &= (\text{T}_{0, i}\times \frac{D(\text{c})}{D_i}+\text{T}_{1,i})\times \frac{\text{f}_i}{\text{f}}+\text{T}_\text{mem,i+1}(\text{c},\text{w}) \nonumber\\ 
    \text{T}_{1,i} &= \text{T}_{\text{BP},i} + \text{T}_{\text{Cache},i} \\
    \text{T}_{0,i} &= \text{T}_i - \text{T}_{1,i} - \text{T}_{\text{mem},i} \nonumber
\end{align}

where D(c) is the dispatch width of core size c. In this equation, the statistics collected over the past interval ($i$) is used as a base for modeling. This simplified analytical model assumes that the execution cycles corresponding to branch prediction (BP) and cache access is not affected considerably by changing the core size, while the rest of the computation cycles is scaled linearly with the dispatch width of the core. Moreover, the memory access time is not affected by the core frequency and estimated as: 

\begin{equation} \label{eq_tmem}
    \text{T}_\text{mem,i+1}(\text{c},\text{w}) = \text{LM}_i(\text{c}, \text{w}) \times \text{L}_{\text{mem}}
\end{equation}

where LM is the number of leading misses collected from ATD and $\text{L}_{\text{mem}}$ is the memory access latency.

Equation \ref{eq_t_model} is used to evaluate QoS as follows: 
\begin{equation}
    \label{eq_qos}
    \text{QoS(c,f,w)}= \begin{cases}
        \text{True}, & \text{if T(c,f,w)$\leq$ T($\text{c}_b, \text{f}_b, \text{w}_b)\times \alpha$}\\
        \text{False} & \text{otherwise}
    \end{cases}
\end{equation}

where $\text{c}_b$, $\text{f}_b$, and $\text{w}_b$ are baseline settings. The $\alpha$ parameter can be used to relax the QoS constraint. However, its value is fixed to 1 in this study.

\subsection{Energy Modeling} \label{subsec_energy_model}
We focus on the effect of a specific target setting on the energy consumption of core and memory accesses for each application. The variation in the energy of other system components is assumed to be negligible in comparison. The core energy is divided into static and dynamic components. The static power is constant and can be measured offline for different core sizes and voltage / frequencies. We assume that the dynamic energy can be estimated online by subtracting the static energy from the measured total core energy. This can be done at specific sampling intervals for each core. Hence, the energy of the upcoming interval ($i+1$) is estimated as: 

\begin{align}\label{eq_e_model}
    \text{E}_{i+1}(\text{c},\text{f},\text{w}) &= [\text{P}_{\text{CoreDyn}}^*(\text{c}) \times \frac{V(\text{f})^2}{{V^*(\text{c})}^2} + \text{P}_{\text{CoreStatic}}(\text{c},\text{f})]\nonumber\\
    &\times \text{T}_{i+1}(\text{c},\text{f},\text{w}) + \text{E}_{\text{mem},i+1}(\text{w}) \\ 
    \text{E}_{\text{mem},i+1}(\text{w}) &= (\text{MA}_i + \text{DM}_i(\text{w})) \times e_{\text{mem}}
\end{align}

where $\text{P}_{\text{CoreDyn}}^*$ is the sampled core dynamic power, $V^*$ is the core voltage during sampling, MA is the total number of memory accesses, DM is the difference in LLC misses for a target allocation compared to the last allocation derived from ATD, and $e_{\text{mem}}$ is the energy of a single memory access. 

\subsection{Overheads}
The overheads imposed by the RM can be divided into three components: i) the area overhead of the proposed extension to the ATD as well as the required support for core adaptation; ii) the instruction overhead of executing the RM algorithm and iii) the time and energy overhead of enforcing the new resource settings. 

As mentioned earlier, the extension to the ATD consists of one extra counter per core size per cache allocation. In this case, we have used three core sizes and 16 possible LLC allocations per core which requires 48 additional counters. Inside each counter, a 27 bit register provides more than enough resolution to count LMs for intervals of 100M instructions. Two more registers are also needed to hold the index of the last LM and the distance of the last OV. We have used 10 bits for instruction index. Therefore, the overall storage requirement is estimated to less than 300 bytes per core, which is almost negligible. Moreover, sending 10 bits of extra information between the core and the ATD only for the memory instructions that miss in the private cache  imposes minor overhead. Nevertheless, these estimates are pessimistic and this technique can be implemented with substantially less overhead after analyzing the sensitivity of the RM to the number of bits in the instruction index and the miss counters. We leave this analysis for future work. 

Adaptive core microarchitecture has been considered in numerous prior work \cite{ElasticCore2018, MLPawareDynIW, DynTunProc,buyuktosunoglu2001circuit} with acceptable hardware overhead. For example, Buyuktosunoglu \textit{et al.} proposed an implementation of adaptive issue queue with less than 3\% gate count overhead~\cite{buyuktosunoglu2001circuit}. A similar estimation was reported in~\cite{MLPawareDynIW} assuming a 32-nm technology node. With the increase in transistor budget in more recent technologies, we assume the required hardware support for the envisioned adaptive architecture to be affordable. 

To evaluate the overheads imposed by the RM algorithm, we measured the instruction count when executing an implementation of the RM in C language. The measurements show 51K, 73K, and 100K executed instructions for two-, four-, and eight-core systems, respectively. In comparison, the algorithm in the previous work \cite{Nejat2019} imposes 18K, 40K, and 67K instruction overhead, respectively. Therefore, for an eight-core system, the overhead is $0.1\%$ assuming 100M instruction interval. 

Finally, to enforce the decisions of the RM, the core-configuration controller and LLC partitioning bit-masks must be updated. The overhead of performing this action is dominated by changing the VF of the core due to its large electrical capacitance. We assume it takes 15 $\mu s$ and consumes 3 $\mu J$ in DVFS overhead as reported in \cite{SangyoungPark2013AccurateMicroprocessors} for the Samsung Exynos 4210. For example, if the clock frequency is set to 2 GHz and the average IPC is 2, this leads to a $0.06\%$ time overhead in a 100M instruction interval which takes 25 ms. The simulation framework explained in Section~\ref{sec_exper_method} accounts for these overheads when invoking the RM during program execution. 

In order to maintain logical correctness during core resizing, the instruction fetch must be halted. It takes a number of cycles equal to the instruction window size divided by the average IPC count to empty the pipeline. Next, the reconfiguration controller deactivates/reactivates sections of ports, buses, and array banks corresponding to the target core size, before restarting the instruction fetch. Finally, the power gating circuit cuts the power to the deactivated sections to reduce static power. Considering the maximum ROB size (see Table~\ref{tbl_base_config}), the run-time overhead of this operation is in the order of a few hundreds of cycle, which is negligible over 100M instruction intervals. Bottomline is that while we take these overheads into account in our results, one can conclude that they are almost negligible considering the size of the intervals.

\section{Experimental Methodology} \label{sec_exper_method}
In this section, after a brief description of the simulation framework, the baseline architecture model is presented. Next, the details of the benchmarks and workloads are discussed, followed by the evaluation metrics.

\subsection{Simulation Framework}
The goal is to evaluate the effect of the RM on performance and energy as applications undergo different phases over their full execution. Therefore, a multi-level framework based on SimPoint analysis~\cite{Sherwood2002AutomaticallyBehavior}, Sniper-7.2 (released 2019)~\cite{Carlson2014AnModels} plus McPAT~\cite{Li2009McPATArchitectures} architectural and power simulations, and an in-house multi-core RM simulator is used. In effect, the proposed ATD-based mechanism to accurately model the impact of MLP is integrated in Sniper while using its ``\textit{ROB}" performance model. 

As the first step, Sniper plus McPAT simulations are performed for representative regions of benchmark phases with 100M warm-up and 100M detailed instruction windows. These simulations are repeated over all possible core configurations, VF settings, and LLC allocations (see Table~\ref{tbl_base_config} for more details). These simulation results are collected in a database for each program phase. 

\begin{figure}[t]
    \centering
    \includegraphics[width=0.8\linewidth]{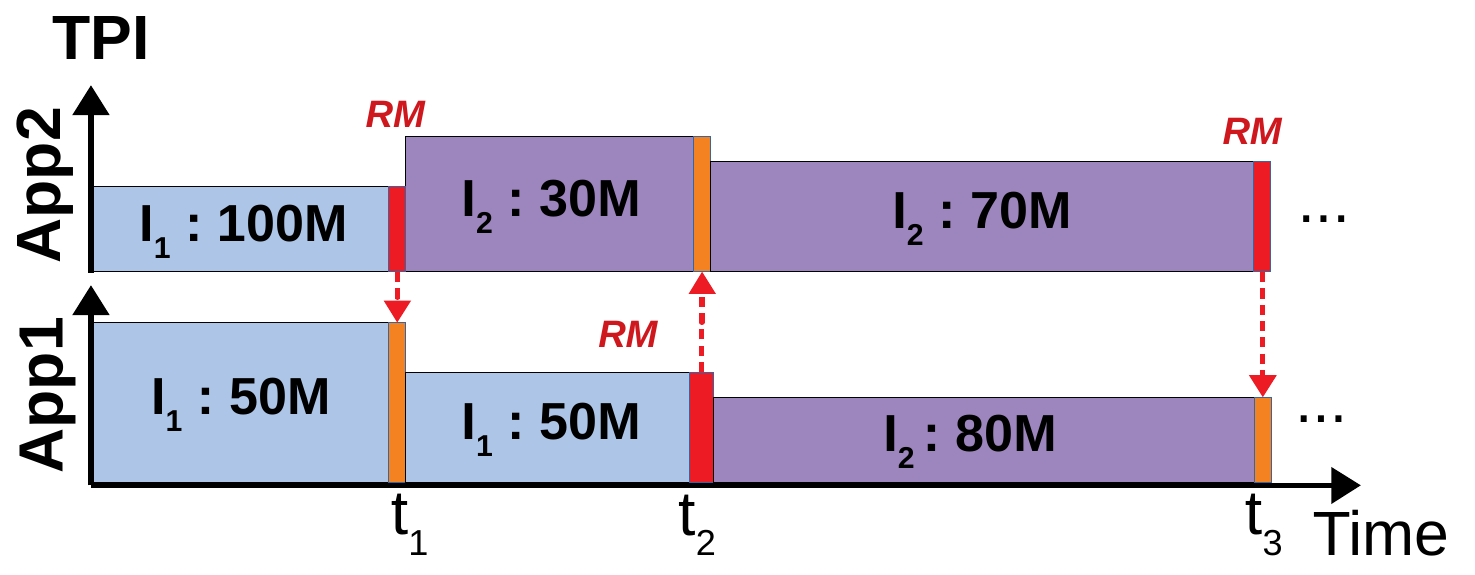}
    \vspace{-5pt}
    \caption{Run time behavior of the RM simulator.}
    \label{fig_rm_sim}
    \vspace{-13pt}
\end{figure}


In the next step, the RM simulator starts by collecting the phase traces generated during SimPoint analysis for each application in the workload. The trace lists the sequence of phase numbers that corresponds to each consecutive program interval. It is used together with the simulation database to generate a proxy of each detailed benchmark execution. Figure~\ref{fig_rm_sim} shows a simple example to explain the mechanics of the RM simulator. The simulation starts in the first program interval (I$_1$) in each application. Using the Time per Instruction (TPI) collected from the simulation database for each application at the baseline setting, the next global event (t$_1$) is found. It is the time when the fastest application finishes one interval. The RM is invoked on the corresponding core to find a new system setting. After updating the statistics for each core with RM overhead, the next global event (t$_2$) is found in a similar fashion. This process continues until the end of simulation. 

\subsection{Base Configuration}

A summary of the base system configurations is reported in Table~\ref{tbl_base_config}. We consider three processor models -- S, M and L -- that model a 2-, 4- and 8-issue processor, respectively. The number of cores depend on the experiments. In this work 2, 4, and 8 core systems are evaluated. 

\newcolumntype{s}{>{\centering \arraybackslash \hsize=0.23\hsize}X}
\newcolumntype{b}{>{\centering \arraybackslash \hsize=0.77\hsize}X}
\vspace{-5pt}
\begin{table}[h]  
  \footnotesize
  \caption{Baseline configuration.}
  \centering
  \setlength\tabcolsep{3pt}
  \begin{tabularx}{\columnwidth}{|s|s|s|s|}
    \hline
    \textbf{Core} & \multicolumn{3}{b|}{out-of-order, branch predictor:Pentium M type}  \\
    \cline{2-4}
     & L & M & S \\ 
    \cline{2-4}
    issue width & 8 & 4 & 2 \\ 
    ROB & 256 & 128 & 64 \\
    RS* & 128 & 64 & 16 \\
    LSQ** & 64 & 32 & 10\\
    \hline
    \hline 
    \textbf{Cache} & \multicolumn{3}{b|}{64B blocks, LRU replacement}\\
    \cline {2-4}
     & L1-I/L1-D & L2 & L3\\ 
     \cline{2-4}
     & Private & Private & Shared \\
    Size & 32 KB & 256 KB & 2 MB $\times$ cores\\
    Associativity & 4 & 8 & 8 $\times$ cores\\
    DVFS domain & core & core & global \\ 
    Allowed Range (per core) & NA & NA & 2 way - 16 way (256KB-4MB)\\ 
    \hline 
    \hline
    \textbf{DRAM} & \multicolumn{3}{b|}{100 ns base latency, contention queue model, 5 GB/s bandwidth per core}\\
    \hline
    \hline 
    \textbf{DVFS} & Core Baseline & Core Range & Global \\ 
    \cline{2-4}
    Frequency & 2 GHz& 1-3.25 GHz & 2 GHz\\
    Voltage & 1 V& 0.8-1.25 V& 1 V\\
    \hline
    \multicolumn{4}{l}{\scriptsize * Reservation Station    ** Load Store Queue} 
  \end{tabularx}
  \label{tbl_base_config}
  \vspace{-10pt}
\end{table}

\subsection{Benchmarks and Workloads} \label{sub_sec_workloads}
The SPEC CPU2006 benchmark suite was selected for this study since it represents a diverse range of workloads. As mentioned earlier in Section~\ref{sec_motivation}, the applications are categorized based on two attributes. First, an application is counted as Cache Sensitive (CS) if the variation in MPKI is greater than $20\%$ when changing the LLC allocation by $\pm$50\% while the MPKI is at least $0.2$ with the baseline allocation i.e. 8-way (2MB). Otherwise, it is considered Cache Insensitive (CI). Next, regarding parallelism, we decided to focus on MLP to simplify the categories. This is based on the observation that MLP has a strong influence on the trade-offs between core configuration and LLC partitioning. Thus, the MLP is estimated on the three core sizes with baseline cache allocation and VF. If the variation in MLP from S to L is greater than $30\%$ of the MLP of the baseline core size (M) while its value is at least 2 on the L core, the application is counted as Parallelism Sensitive (PS); otherwise it is Parallelism Insensitive (PI). 

Table~\ref{tbl_categories} presents applications that belong to each category. For two of the SPEC CPU2006 applications, namely \textit{calculix} and \textit{milc}, the Sniper simulations did not finish properly in some benchmark phases. Therefore, we excluded them from this study.

\newcolumntype{s}{>{\centering \arraybackslash \hsize=0.13\hsize}X}
\newcolumntype{b}{>{\centering \arraybackslash \hsize=0.87\hsize}X}

\begin{table}[h]  
  \footnotesize
  \caption{Application categories.}
  \centering
  \setlength\tabcolsep{1.5pt}
  \begin{tabularx}{\columnwidth}{|s|b|}
    \hline
    \textbf{Category} & \textbf{Applications} \\
    \hline
    CS-PS & tonto, mcf, omnetpp, soplex, sphinx3\\
    \hline
    CS-PI & bzip2, gcc, gobmk, gromacs, h264ref, hmmer, xalancbmk\\
    \hline 
    CI-PS & namd, zeusmp, GemsFDTD, bwaves, leslie3d, libquantum, wrf\\
    \hline 
    CI-PI & cactusADM, dealII, gamess, perlbench, povray, sjeng, astar, lbm\\
    \hline 
     
  \end{tabularx}
  \label{tbl_categories}
\end{table}

In Section~\ref{sec_motivation} all possible mixes of application categories are analyzed in a two-core workload. This analysis provides interesting insights about the workload scenarios where the proposed RM improves the energy savings and where its effectiveness is limited. Hence, we use the same four scenarios (bounded rectangles in Table~\ref{tbl_categories}) to evaluate the energy savings. But, we extend each scenario to four and eight core workloads as follows. For the first half of the cores, a benchmark application is randomly selected from a category corresponding to \textit{App1} in Table~\ref{tbl_categories}. Similarly, the applications for the second half of the cores are selected according to \textit{App2}. For example, to create Scenario~1, the first half can be from any category as long as the second half is selected from CS-PS. Additionally, the second half can be CS-PI if the first half is CI-PS. The Python function \texttt{random.choice} is used to select benchmark applications.The processes is repeated until each application is selected at least once over all workloads. 

\subsection{Evaluation Metrics} \label{subsec_eval_metrics}

The total number of instructions varies significantly across benchmark applications. Therefore, in order to have a fair comparison, the simulations are run until each application in the workload has executed at least once. This is 4146B instructions as it is the (dynamic) instruction count  of the longest application among the benchmarks. 
Each application is re-stared until the end of simulation. 

\subsubsection{Energy Savings}

The energy consumption measured for each case is calculated as the sum of core and dynamic energy of memory for every application until it has executed 4146B instructions, plus the un-core (LLC and network-on-chip) energy until the end of simulation. This value is compared for each RM to an idle RM that keeps the baseline system setting until the end of simulation. The same three RMs are evaluated as mentioned in Section~\ref{sec_motivation}. 

\subsubsection{QoS} \label{subsec_method_qos}
When the RM is invoked at every execution interval, it attempts to find a resource setting that satisfies QoS according to Equation~\ref{eq_qos}. This requires performance modeling for both the target setting and the baseline setting. However, due to modeling error, the RM may select a setting that violates QoS for the next interval. 

Therefore, an extensive evaluation is performed to estimate the probability and expected value of QoS violations over all benchmark applications as follows. QoS is violated over the next interval if all these conditions are met: 
\begin{enumerate}
    \item For actual values: $\text{T}^{\text{Act.}}_{i+1}$(\textit{Target}) $> \text{T}^{\text{Act.}}_{i+1}$(\textit{Base})
    \item For predicted values: $\text{T}_{i+1}$(\textit{Target}) $\leq \text{T}_{i+1}$(\textit{Base})
    \item The \textit{Target} setting is selected by the RM
\end{enumerate}  

The modeling of interval $i+1$ is performed using the statistics collected at interval $i$. Therefore, the error depends on the current setting during interval $i$. The probability of QoS violation is evaluated by iterating over all phases of all applications, all possible current settings, and all possible target settings and checking the above conditions. We assume equal probability for any current setting and selection of any target setting. The phase weights generated by SimPoint is used as the probability of each program phase. 

In cases of QoS violation, the amount of violation is calculated as: 

\begin{equation} \label{eq_violation}
    \text{Violation} = \frac{\text{T}^{\text{Act.}}(\text{Target}) - \text{T}^{\text{Act.}}(\text{Base})} {\text{T}^{\text{Act.}}(\text{Base})}
\end{equation}

The same probabilities are used to calculate the expected value and standard deviation of violations over all violating cases.

\section{Evaluation Results} \label{sec_exper_res}
The experimental results are reported in this section. First, the energy savings are analyzed. Next, we study how accurately QoS is tracked. Finally, the effect of modeling error on energy savings is studied. 

\begin{figure*}[ht]
    \centering
    \includegraphics[width=\linewidth]{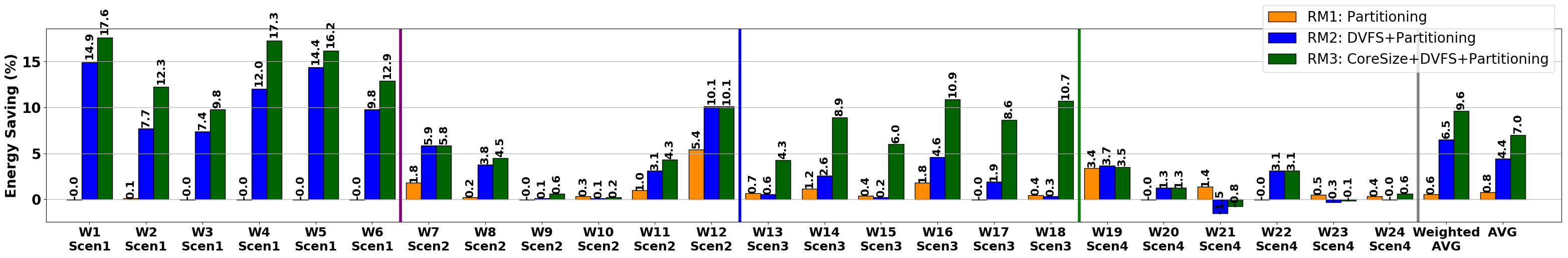}
    \includegraphics[width=\linewidth]{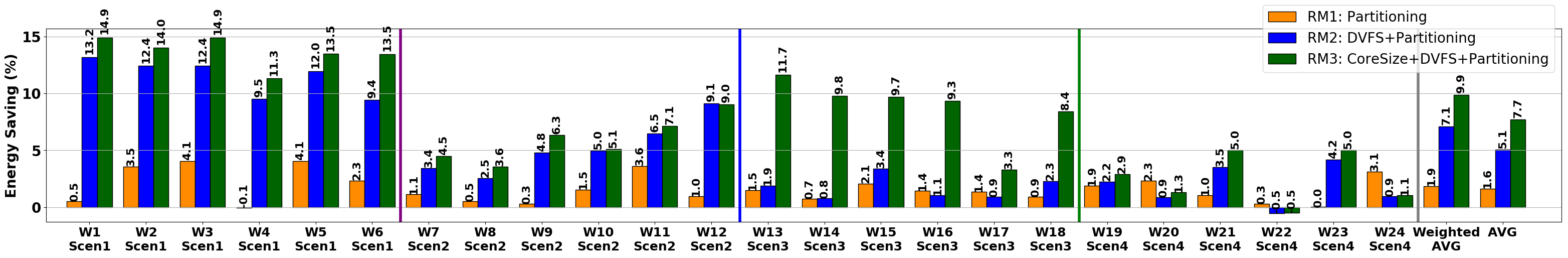}
    
    \caption{Energy savings with three different RMs on 4 core (top) and 8 core (bottom) workloads.}
    \label{fig_eimp_results}
    \vspace{-10pt}
\end{figure*}

\subsection{Energy Savings}
To evaluate the proposed scheme, we perform experiments on six 4-core and six 8-core workloads generated according to Section~\ref{sub_sec_workloads} for each of the four scenarios described in Section~\ref{sec_motivation}. The resulting energy savings are shown in Figure~\ref{fig_eimp_results}. From left to right, the first bar represents RM1 that only performs LLC partitioning. The second bar corresponds to RM2 that coordinates per-core DVFS with LLC partitioning. Finally, the proposed scheme (RM3) corresponds to the third bar. These results include the effect of performance and energy modeling error and the overheads. 

The results in Figure~\ref{fig_eimp_results} are grouped based on the workload scenarios mentioned in the labels for each set of bars. However, these scenarios do not occur with equal probability. Therefore we use the probabilities mentioned in Figure~\ref{fig_scenarios_table} to calculate the average energy savings. These probabilities were calculated using the distribution of benchmarks in Table~\ref{tbl_categories}. Hence, we have used 47\%, 22.1\%, 22.1\%, and 8.8\% as the weights for scenarios 1 to 4 respectively. The normal average results are also added. 

Recall that in the first scenario, RM3 is expected to achieve additional energy saving compared to RM2. The results for Scenario~1 shows up to $17.6\%$ energy saving with RM3 compared to $14.9\%$ with RM2 in the same workload (\textit{4Core-W1}). The relative improvement is greater in other workloads such as \textit{4Core-W2} where RM3 achieves $60\%$ higher energy saving compared to RM2. The trend is similar for 8 core workloads. 

In the second scenario, the energy savings are expected to be smaller compared to the first scenario. The results show smaller difference between RM2 and RM3 as expected. In the case of \textit{4Core-W9} and \textit{4Core-W10} the energy savings are negligible. The former is a mix of three \textit{xalancbmk} and one \textit{hmmer} while the latter is a mix of \textit{gcc}, \textit{gromacs}, \textit{h264ref}, and \textit{gobmk}. All these benchmarks belong to the CS-PI category. While all these applications are counted as cache sensitive, except \textit{xalancbmk} and \textit{h264ref}, they do not benefit considerably by an increase in their baseline allocation. But, a reduction in the allocations leads to a substantial increase in the MPKI. In other words, the baseline equal distribution is already an optimum allocation for these workloads. 

The results for the third scenario shows a significant improvement with RM3 compared to the other two RMs. As explained in Section~\ref{sec_motivation}, in this scenario RM1 and RM2 are not effective since applications are not sensitive to their LLC allocation. RM3 on the other hand saves energy by making a trade-off between core size and VF. 

As expected, the general trend in the forth scenario shows negligible and small improvements with all three RMs. In some cases (\textit{4Core-W21} and \textit{8Core-W22}) there are even small increases in the system energy due to modeling errors. In the case of \textit{8Core-W20} and \textit{8Core-W24}, these errors result in a lower energy saving with RM2 and RM3 compared to RM1 that does not change the core size and VF. The effect of modeling accuracy on energy savings is further discussed in the following subsection where the results using perfect models are also analyzed.

The analysis based on workload scenarios clearly shows the advantage and limitations of the proposed RM. If the workload can be classified into the first or third scenarios, the system energy can be reduced substantially while respecting the performance constraints of every application. Focusing on the simulation results (4 and 8 core) for Scenario~1 (47\% probability), RM3 can save on average about 14\% of energy compared to 11\% with RM2. In Scenario~3 (22\% probability), the average energy saving is about 8.5\% with RM3 compared to 1.7\% with RM2. On the other hand, if the workload is classified into the second or forth scenarios (31\% probability), lower average energy savings which is close to the previous work should be expected from the proposed scheme.

\subsection{QoS Evaluation}
This paper presents a modeling framework that estimates the impact of MLP on performance to accurately track QoS targets. To evaluate the effectiveness of this framework (Model3), we compare it with the modeling approach used in the previous work \cite{Nejat2019} (Model2). In that approach the MLP is assumed to be constant across different resource settings. Hence the value measured over the past interval is used across all target settings in the next interval. Additionally, we compare against a simple model that multiplies the total number of cache misses by the memory latency to calculate the memory access time (Model1). 

\begin{figure}[t]
    \centering
    \includegraphics[width=0.75\linewidth]{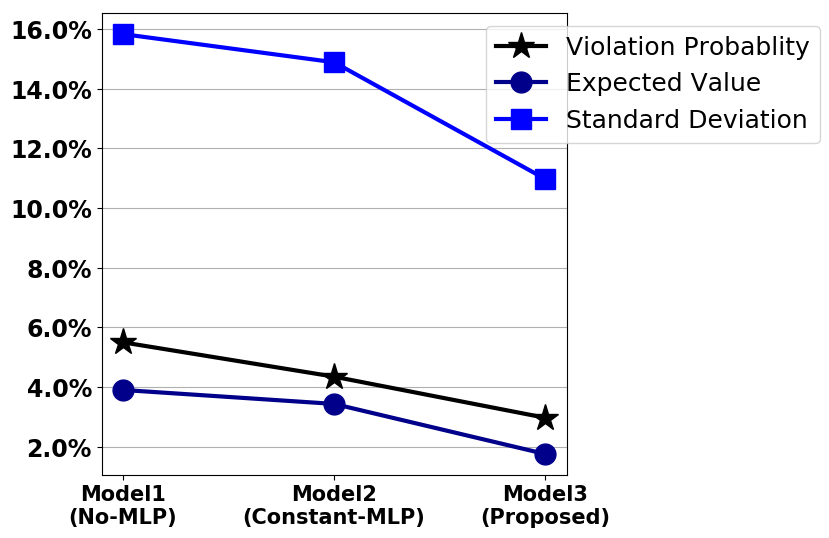}
    \caption{Probability of QoS violation plus expected value and standard deviation of violations using different models.}
    \label{fig_qos_stats}
    \vspace{-10pt}
\end{figure}

\begin{figure}[t]
    \centering
    \includegraphics[width=0.85\linewidth]{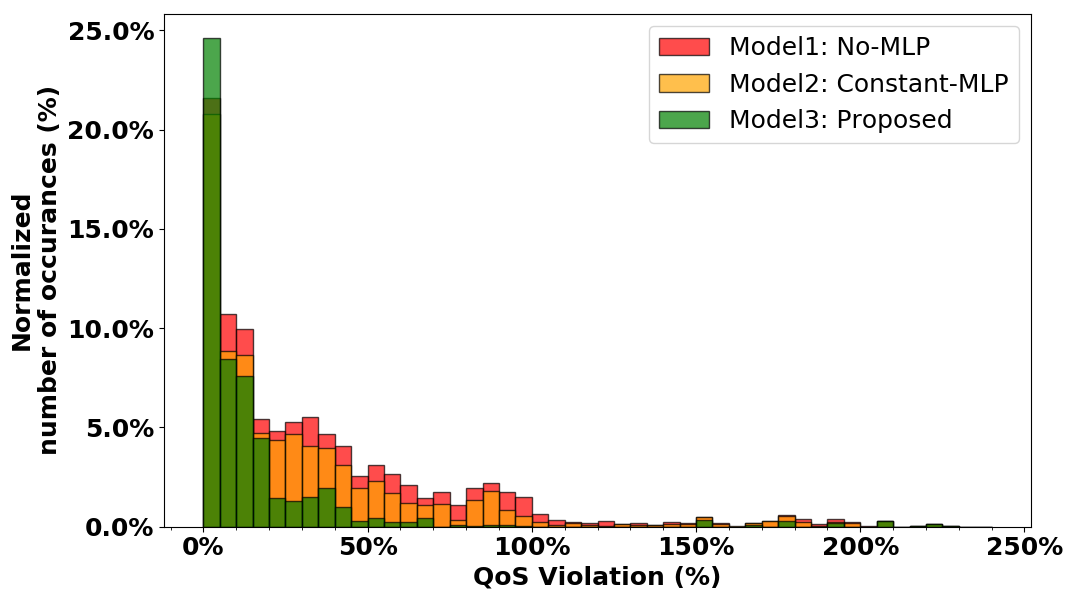}
    \caption{Distribution of QoS violations using different performance models.}
    \label{fig_qos_hist}
    \vspace{-10pt}
\end{figure}

Figure~\ref{fig_qos_stats} shows the probability of QoS violation in an execution interval for the three models. It also shows the expected value and standard deviation of violations calculated by Equation \ref{eq_violation}. The figure shows that Model3 reduces QoS violation probability by 46\% and 32\% compared to Model1 and Model2 respectively. Furthermore, the expected value of violations and the standard deviation also reduces substantially by 49\% and 26\%, respectively, compared to Model2. 

A more detailed view of the QoS violations is illustrated in Figure~\ref{fig_qos_hist}. Here, the X-axis shows the amount of violation, while the Y-axis represents the number of occurrences normalized to the maximum number of violations across the three models. According to this figure, the proposed model has slightly more number of violations in the range of 5\%. But, the total number of violations is substantially smaller compared to Model~1 and Model~2. More specifically, the violations with a larger value -- tail latency -- has reduced significantly. 

\begin{figure*}[ht]
    \centering
    \includegraphics[width=\linewidth]{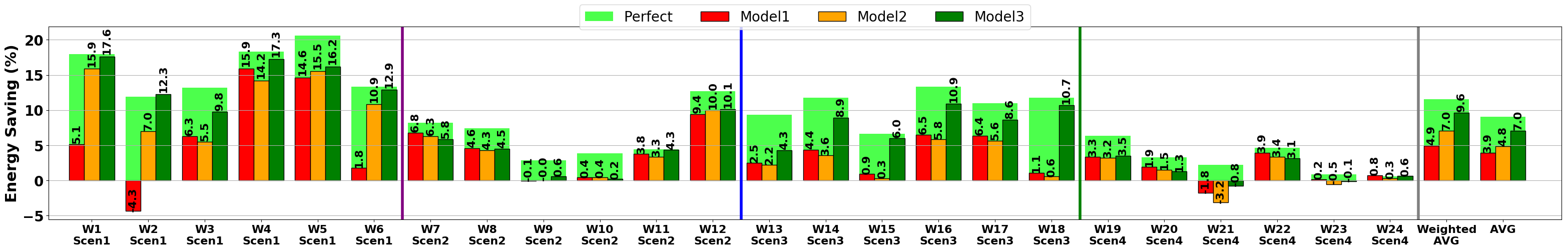}
    \includegraphics[width=\linewidth]{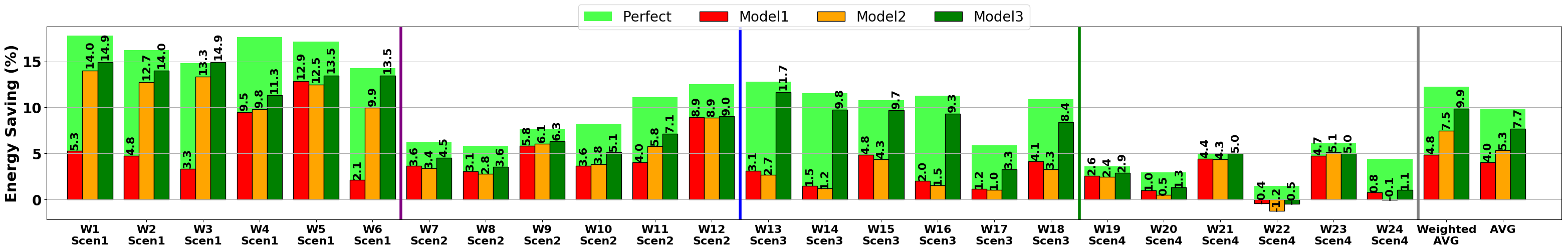}
    \caption{Comparison of the energy savings achieved with the proposed RM on 4-core (top) and 8-core (bottom) workloads when using different performance models.}
    \label{fig_3model_eimp}
    \vspace{-10pt}
\end{figure*}

\subsection{Effect of Modeling Error}
The modeling error affects the achieved energy savings. Figure~\ref{fig_3model_eimp} shows the energy savings in 4-core (top) and 8-core (bottom) workloads using the proposed modeling framework (Model3) as well as the other two models described earlier. The light green box shows the energy savings achieved if perfect performance and energy models were used and the program phase during next interval is always predicted accurately. The results in this figure are derived using the proposed RM3. By comparing the energy savings of each model to the perfect results, we observe that the proposed modeling framework achieves higher energy saving that is closer to the perfect results.

\section{Related Work} \label{sec_related}


DVFS is an effective technique to make trade-off between performance and energy and is widely used to save energy under QoS constraints for example in \cite{MOGHADDAM2017, Suh2015DynamicProcessors,Kasture2015Rubik:Systems,Choi2002Frame-basedDecoder}. During program execution, DVFS in isolation is only effective if some reduction of performance is acceptable, as shown in \cite{MOGHADDAM2017, Nejat2019}. In studies like \cite{Suh2015DynamicProcessors} where the QoS target is expressed as a fixed instruction per second rate during program execution, DVFS has limited efficiency. Such controller uses a higher VF during memory intensive phases compared to compute intensive phases to maintain a stable instruction per second rate. While core frequency has small effect during memory intensive phases, increasing cache resources or MLP may lead to significant improvement in performance and energy efficiency. 

Partitioning of the shared LLC has been used in prior art \cite{PARTIES2019, Funaro2016Ginseng:Allocation, Lo2015Heracles:Scale, Kasture2014Ubik, METE2011, Moreto2009FlexDCP:Architectures,Takagi2009CooperativeMultiprocessors, Fu2011Cache-AwareSystems, Nejat2019} for applications with QoS constraints. However, these works do not consider adaptation of core microarchitecture. Such adaptive architectures provide a means to trade core energy with ILP and MLP. The amount of MLP considerably affects the performance cost of each individual memory access. The cache partitioning algorithm must take this into account as argued in \cite{MLP-Partitioning2007}. This is especially important when applications have performance constraints. 

Dynamic adaption of core micro-architecture has been the focus of numerous studies \cite{ElasticCore2018, PhBasedDynIWOpt, MLPawareDynIW,Petoumenos2010, SmthgOldSmthngNew, DynTunProc,Buyuktosunoglu2000}. They have demonstrated the potential of this technique in improving the energy efficiency of the core. 
However, these works do not consider energy optimization under application QoS constraints. These constraints are considered in the context of core reconfiguration for multi-media applications in \cite{Hughes2001SavingApplications,Sasanka2002} and in a more general form in \cite{Pothukuchi2016UsingArchitectures, Zhou2016CASH}. Pothukuchi \textit{et al.} \cite{Pothukuchi2016UsingArchitectures} designed a controller that takes multiple inputs including core VF and micro-architectural parameters and tracks multiple targets such as energy and performance constraints. Zhou \textit{et al.} \cite{Zhou2016CASH} propose a flexible core architecture by combining fixed slices and L2 cache banks. In their proposal, a controller monitors application QoS during run-time and adapts the configuration to meet QoS targets with minimum service cost. However, none of these works consdier partitioning of a shared LLC among multiple applications. 

Despite its significant potential, to the best of our knowledge, no prior art has studied the combined management of core micro-architectural resources, DVFS, and LLC partitioning under QoS constraints for all applications. This enables trading resources between applications based on their online characteristics to reduce core and memory access energy without any performance degradation. 

An accurate estimation of performance is needed to predict if a particular resource allocation leads to QoS violation. Previous studies demonstrate the importance of MLP in performance estimation \cite{LeadingLoads2014, Miftakhutdinov2012}. They propose to count only the leading loads in a series of overlapping memory accesses. Spiliopoulos \textit{et al.} \cite{spiliopoulos2016unified} extended this approach to estimate leading loads across different cache allocations. However, these proposals cannot predict the effect of core adaptation on MLP. Therefore, this work propose a low-cost hardware extension to address this problem.

\section{Concluding Remarks} \label{sec_conclude}
It is important to support multiple applications with QoS targets in a multi-core system. But, respecting the performance constraints of all applications complicates the management of hardware resources and hinders energy optimization. Prior work has proposed coordinated management of per-core DVFS and cache partitioning to address this problem. However, its scope for energy optimization is limited due to the quadratic energy cost of DVFS and its futility over memory access time. 

This work propose a new resource-management framework that simultaneously throttles processor micro-architecture resources, DVFS, and partitioning of the shared cache. By exploiting the new resource trade-offs enabled by ILP and MLP exploitation, the proposed framework achieves higher energy savings compared to the previous work, over the majority of possible workload mixes. Our evaluations shows that the energy of 4-core and 8-core systems can be reduced by up to 18\% and on average 10\%. Furthermore, we propose a low-overhead hardware mechanism  to estimate MLP across different core configurations and cache allocations. It enables analytical performance modeling with improved accuracy and does not depend on any prior information about the applications. This framework reduces the probability and expected value of QoS violations by 32\% and 49\% respectively, compared to previous approaches.

\section*{Acknowledgments} 
This research has been funded by the European Research Council, under the MECCA project, contract number ERC-2013-AdG 340328.

\bibliographystyle{IEEEtran}
\bibliography{references.bib}

\end{document}